\documentclass[conference]{IEEEtran}
\usepackage{fancyhdr}

\ifCLASSOPTIONcompsoc
  \usepackage[nocompress]{cite}
\else
  \usepackage{cite}
\fi
\usepackage[hidelinks]{hyperref}

\usepackage{algorithm}
\usepackage{algpseudocode}
\usepackage{amsmath}
\usepackage{booktabs} %
\usepackage{url}

\usepackage{listings}
\makeatletter
\let\@ORGmakecaption\@makecaption
\long\def\@makecaption#1#2{\@ORGmakecaption{#1}{#2}\vskip\belowcaptionskip\relax}
\makeatother
\usepackage{float}

\usepackage{pgfplots}
\pgfplotsset{compat=1.9}
\usepgfplotslibrary{groupplots}

\usepackage{tikz}
\usetikzlibrary{positioning,shadows,arrows,trees,shapes,fit}

\title{Static Generation of Efficient OpenMP Offload Data Mappings}

\author{
  \IEEEauthorblockN{Luke Marzen}
  \IEEEauthorblockA{\textit{Department of Computer Science} \\
                    \textit{Iowa State University}          \\
                    Ames, IA, USA                           \\
                    ljmarzen@iastate.edu }
  \and
  \IEEEauthorblockN{Akash Dutta}
  \IEEEauthorblockA{\textit{Department of Computer Science} \\
                    \textit{Iowa State University}          \\
                    Ames, IA, USA                           \\
                    adutta@iastate.edu}
  \and
  \IEEEauthorblockN{Ali Jannesari}
  \IEEEauthorblockA{\textit{Department of Computer Science} \\
                    \textit{Iowa State University}          \\
                    Ames, IA, USA                           \\
                    jannesar@iastate.edu}
}

\begin{document}

\maketitle

\thispagestyle{fancy}
\lhead{}
\rhead{}
\chead{}
\lfoot{\footnotesize{
SC24, November 17-22, 2024, Atlanta, Georgia, USA
\newline 979-8-3503-5291-7/24/\$31.00 \copyright 2024 IEEE}}
\rfoot{}
\cfoot{}
\renewcommand{\headrulewidth}{0pt}
\renewcommand{\footrulewidth}{0pt}

\begin{abstract}

Increasing heterogeneity in HPC architectures and compiler advancements have led to OpenMP being frequently used to enable computations on heterogeneous devices.
However, the efficient movement of data on heterogeneous computing platforms is crucial for achieving high utilization. 
Programmers must explicitly map data between the host and connected accelerator devices to achieve efficient data movement.
Ensuring efficient data transfer requires programmers to reason about complex data flow. 
This can be a laborious and error-prone process since the programmer must keep a mental model of data validity and lifetime spanning multiple data environments. 
We present a static analysis tool, {\tt OMPDart} (\underline{O}pen\underline{MP} \underline{Da}ta \underline{R}eduction \underline{T}ool), for OpenMP programs that models data dependencies between host and device regions and applies source code transformations to achieve efficient data transfer.
Our evaluations on \textit{nine} HPC benchmarks demonstrate that {\tt OMPDart} is capable of generating effective data mapping constructs that substantially reduce data transfer between host and device.

\end{abstract}

\begin{IEEEkeywords}
High performance computing, Parallel programming, Data transfer, Static analysis, GPU, Performance optimization, Heterogeneous computing, OpenMP
\end{IEEEkeywords}

\section{Introduction}

High-performance heterogeneous computing architectures have increasingly deployed accelerators such as GPUs.
In recent years, the Top500 List \cite{top500} has been dominated by supercomputers leveraging GPU accelerators to achieve higher performance and efficiency.
This paradigm shift offers unprecedented computational power, but it also brings forth challenges in efficiently managing data movement across diverse computing resources, as each connected device typically has its own distinct memory space. 
In this context, the efficient movement of data is paramount to achieving high performance, as redundant data transfers introduce overheads in terms of execution time and energy.

OpenMP is a portable parallel programming model that has been primarily used for easily parallelizing serial code.
Over the past few years, it has been increasingly used to offload computations to accelerators.
The implicit data-mapping rules that apply by default in OpenMP offload applications often lead to redundant data transfers.
To address these challenges, programmers are required to explicitly define data mappings between the host (CPU) and connected accelerator devices (typically GPUs) using OpenMP's \texttt{target data} and \texttt{target update} constructs.
Given that inefficient data transfer can dominate execution time, reducing data transfer while maintaining correctness stands out as a paramount optimization goal for developers.
However, the complexity of the OpenMP specifications in combination with the necessity to reason about data flow across multiple memory spaces for each variable make this a nontrivial and time-consuming task.

In this work, we present static data-flow analysis techniques that include interprocedural and context-sensitive approaches to identify data dependencies between memory spaces.
We introduce {\tt OMPDart}, a static analysis tool for OpenMP programs designed to automate the data mapping process by modeling data dependencies between host and device regions and inserting data-mapping constructs in the source code to resolve them.
Our techniques are based on a hybrid code representation that combines an Abstract Syntax Tree (AST) and a Control Flow Graph (CFG).
Nine HPC benchmarks are used to evaluate the tool by comparing the performance of the tool-defined mappings against an expert hand-optimized implementation.
We show that {\tt OMPDart} can correctly generate data mappings that perform substantially better than default data mappings and on par with or, in some cases, better than the expert-defined mappings.  

In this paper, we make the following contributions:
\begin{itemize}
    \item Introduced static data-flow analysis techniques for identifying data dependencies between memory spaces.
    \item Implemented a hybrid AST-CFG representation to perform interprocedural and data-flow analysis to appropriately map data to memory spaces.
    \item Leveraged the Clang/LLVM toolchain to implement the proposed techniques in a source code transformation tool that enables the automatic generation of OpenMP data mapping directives.
    \item Experimentally evaluated our tool on nine HPC benchmarks and compared it against expert-defined data mappings, showing competitive performance with the expert implementations and a significant reduction in data transfer over the implicit OpenMP data mapping rules.
    \item Improved the performance of the OpenMP offload version of LULESH 2.0 in HeCBench by using {\tt OMPDart} to identify alternative data mappings, resulting in an 85\% reduction in data transfer and a 1.6$\times$ speedup over expert-defined mappings on a medium input size.
\end{itemize}

The remainder of this paper is organized as follows.
Section \ref{sec:background} provides background on OpenMP offloading and static data-flow analysis.
Section \ref{sec:motivation} provides a motivating example, followed by our proposed approach in Section \ref{sec:approach} and experimental setup and evaluations in Sections \ref{sec:benchmarks} and \ref{sec:results}.
We discuss our approach and results in Section \ref{sec:discussion} and conclude the paper with Section \ref{sec:conclusion}.

\section{Background} \label{sec:background}

\subsection{OpenMP}
{\tt OpenMP} is a widely used shared-memory programming model for parallelizing serial code.
In its simplest form, OpenMP requires users to simply insert {\tt pragma}s before appropriate code blocks (such as loops) to distribute the workload between multiple processors.
With the 4.0 specifications \cite{openmpSpec40}, OpenMP introduced a set of directives to enable compilers to offload work to devices such as GPUs.
This helped abate dependence on vendor-specific compiler infrastructures and ease the use of GPUs for general-purpose computing.
Offloading with OpenMP consists of two high level steps:  (i) transfer and map data between host and device, (ii) offload computations to the device.
Offloading computations to GPUs can potentially provide significant performance and energy efficiency improvements over traditional CPU-based multi-core processing. 
OpenMP offload directives enable effective exploitation of coarse-grained and scalable parallelism on GPUs.
With the drastic improvement in GPU hardware, data mapping between host and device is now one of the primary bottlenecks in OpenMP offload computations.
To this end, OpenMP introduced the {\tt data map} clauses to allow programmers more fine-grained control over data transfer between host and device.
However, this is not always straightforward to accomplish, especially without an in-depth understanding of an application or its runtime behavior.
Our proposed tool aims to automate the selection of such clauses and help programmers improve performance by reducing data transfer overheads.

\subsection{Static Data-Flow Analysis}

\textit{Flow-sensitive analysis} considers statement order in a program and tracks data constraints at each point of a program. 
For this, \textit{Control Flow Graphs} (CFGs) are typically used. 
The nodes of a CFG consist of basic blocks and its directed edges represent the control flow paths between different basic blocks.

\textit{Path-sensitive analysis} extends \textit{flow-sensitive analysis} by considering all possible execution paths of a program. 
While \textit{path-sensitive analysis} is more precise, it increases the time and space complexity since the CFG grows exponentially.

\textit{Context-sensitive analysis}, sometimes referred to as \textit{polyvariant analysis}, is a type of \textit{interprocedural analysis} concerned with understanding the influence of calling contexts, i.e. the program state at function entry. 

\textit{Liveness analysis}, also known as \textit{live-variable analysis}, is a type of data flow analysis that is used to determine which variables are \textit{live} at a given point in the program's execution. 
A variable is considered \textit{live} if it holds a value that may be used or referenced at some point in the program's future execution. 
Common uses for \textit{liveness analysis} include determining efficient register allocation during compilation and facilitating dead-code elimination.

\section{Motivation} \label{sec:motivation}

This section will review common programming patterns that result in redundant data transfer.
We will also demonstrate how programmers can easily introduce bugs while trying to minimize communication overhead, highlighting the error-prone nature of manually performing this endeavor.

Listing \ref{lst:motivation_1} depicts a kernel called repeatedly inside a loop.
Since no explicit data mappings are present, the implicit data-mapping rules are applied.
Relying on the implicit data-mapping rules results in data being copied to the device on entry of an OpenMP target region and copied from the device on exit.
In Listing \ref{lst:motivation_1}, this means coping array \texttt{a} to and from the device every iteration of the outer loop.

\begin{lstlisting}[
  label=lst:motivation_1,
  caption={Kernel nested inside a loop creates redundant data transfer each iteration of the outer loop.},
  captionpos=b,
  language=C,
  numbers=left,
  stepnumber=1,
  tabsize=2,
  showspaces=false,
  showstringspaces=false,
  basicstyle=\ttfamily\footnotesize,
  xleftmargin=20pt,
  float=h,  %
  aboveskip=-\parskip,
  belowskip=-\parskip,
]
int a[N] = {};
for (int i = 0; i < N; ++i) {
    #pragma omp target
    for (int j = 0; j < N; ++j) {
        a[j] += j;
    }
}    
\end{lstlisting}
\vspace{-3mm}
\begin{lstlisting}[
  label=lst:motivation_2,
  caption={Redundant data transfer between kernels due to reliance on implicit data-mapping rules.},
  captionpos=b,
  language=C,
  numbers=left,
  stepnumber=1,
  tabsize=2,
  showspaces=false,
  showstringspaces=false,
  basicstyle=\ttfamily\footnotesize,
  xleftmargin=20pt,
  float=h,  %
  aboveskip=-\parskip,
  belowskip=-\parskip,
]
int a[N] = {};
#pragma omp target
for (int i = 0; i < N; ++i) {
    a[i] += i;
}

#pragma omp target
for (int i = 0; i < N; ++i) {
    a[i] *= i;
}
\end{lstlisting}

Listing \ref{lst:motivation_2} depicts redundant data transfer between kernels.
After execution of the first \texttt{for} loop, the default behavior is to copy array \texttt{a} back to the host then immediately back to the target, even though no data has been written to \texttt{a} in between the execution of these kernels.
To reduce the communication overhead in Listings \ref{lst:motivation_1} and \ref{lst:motivation_2}, programmers must explicitly map each variable between the host and accelerator devices, using OpenMP's \texttt{target data} and \texttt{target update} constructs.
These simple examples represent patterns of redundant data transfer found in real world applications.

\begin{lstlisting}[
  label=lst:motivation_3,
  caption={Incorrect usage of the map construct.},
  captionpos=b,
  language=C,
  numbers=left,
  stepnumber=1,
  tabsize=2,
  showspaces=false,
  showstringspaces=false,
  basicstyle=\ttfamily\footnotesize,
  xleftmargin=20pt,
  float,  %
  belowskip=-\floatsep,
]
int a[N] = {}, sum = 0;
#pragma omp target data map(a)
for (int i = 0; i < M; ++i) {

    // incorrect mapping
    #pragma omp target map(from:a)
    for (int j = 0; j < N; ++j) {
      a[j] += j;
    }
    
    for (int j = 0; j < N; ++j) {
      sum += a[j];
    }
}    
\end{lstlisting}

Listing \ref{lst:motivation_3} shows an incorrect data mapping. 
In this example the programmer intends for the array to be transferred from the device back to the host to be summed each iteration.
However, the programmer noticed that the implicit data mappings would introduce data movement in both directions each iteration of the outer loop.
In an attempt to eliminate the redundant data transfer, the programmer created a data region that moves array \texttt{a} before and after executing the outer loop, then created a second mapping on the first inner loop to transfer the data back to the host for summation.
The actual behavior of this code is to only transfer the data before and after executing the outer loop.
This is because OpenMP 5.2 uses a reference count mechanism to decide when to copy data to and from a device map environment \cite{openmpSpec52}.
The reference count is incremented every time a new device map environment is created and decremented when exiting a region with the \texttt{from} or \texttt{release} map-type.
Data is only actually copied from the device to the host when the reference count is decremented to zero.
To get the expected behavior, the programmer should remove the inner loop data mapping and instead insert an \texttt{update from} clause directly after the first inner loop.
It is clear from this initially benign example that the data mapping rules are more complicated than may initially meet the eye and necessitate a deep understanding to use efficiently and correctly.
The complexity of this problem is exacerbated when considering the interactions of global variables and data mapping regions that map across multiple functions.
This is obfuscated further when accessing the data mapping of many variables with less trivial control flow.
Automating this task has the potential to improve the performance of applications on heterogeneous computing platforms while simultaneously reducing the time and effort required to create these programs.
To the best of our knowledge, no prior works detail context- and flow-sensitive techniques that can correctly generate efficient OpenMP data mapping constructs for complex real-world programs.

\begin{figure*}[htbp]
    \centering
    \includegraphics[ width=0.7\textwidth]{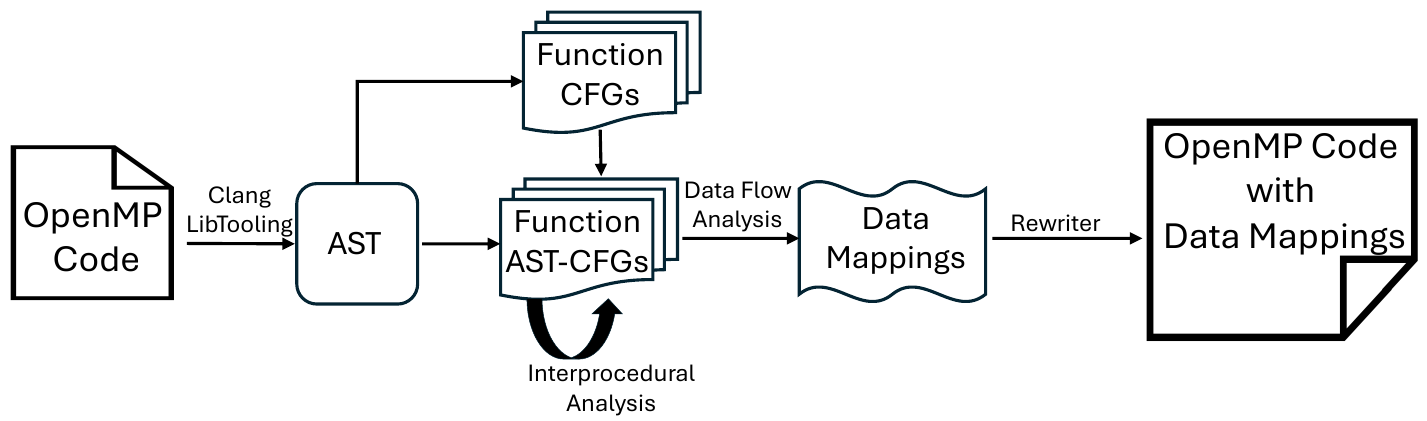}
    \caption{Workflow used by {\tt OMPDart} for identifying data dependencies and transforming source code for C and C++ OpenMP GPU offloading applications.}
    \label{fig:pipeline}
    \vspace{-5mm}
\end{figure*}

\section{Approach} \label{sec:approach}

Our approach begins by leveraging the LLVM toolchain to parse the input source code and generate an AST.
We then transform the AST into CFGs for each function and construct a hybrid data structure that integrates the AST of a function with the corresponding CFG, which we refer to as an \textit{AST-CFG}.
The AST is analyzed to identify operations that perform memory accesses.
An interprocedural analysis is conducted on the AST-CFGs to augment each graph with the side effects of function calls.
These side effects encompass any memory accesses to variables outside the local scope of a function, such as global variables and variables passed by reference.
Next, we traverse the AST-CFG of each function to identify potential data dependencies between the host and device memory spaces and make determinations about how to most efficiently satisfy the dependencies using OpenMP data mapping directives.
Finally, the list of data mapping directives, along with the statements and variables they correspond to, are passed to the \textit{Rewriter} stage (Section \ref{sec:rewriter}), which is responsible for consolidating and determining the final insertion location of each directive and outputting the transformed source code. 
An overview of our workflow is shown in Figure \ref{fig:pipeline}.

\subsection{Input} \label{sec:input}

The expected input is valid C/C++ source code with OpenMP offloading directives. 
This code should not include any instances of \texttt{target data} or \texttt{target update} directives.
{\tt OMPDart} aims to ease programmer effort and automatically insert the appropriate OpenMP data directives.
It is important to acknowledge that in some cases a programmer may deliberately intend to use stale data. 
In such scenarios, our tool would be unfit as it lacks the capability to discern the programmer's intentions.
However, the tool's output could serve as a helpful starting point that the programmer can manually modify to achieve their desired behavior.

\subsection{Parsing the Abstract Syntax Tree}
\begin{table}
\centering
\caption{AST Nodes recognized as offload kernels.}
\begin{tabular}{p{0.51\linewidth}p{0.4\linewidth}}
\hline
\textbf{Clang AST Node} & \textbf{OpenMP Directive}\\
\hline
\textit{OMPTargetDirective} & \texttt{omp target}\\
\textit{OMPTargetParallelDirective} & \texttt{omp target parallel}\\
\textit{OMPTargetParallelForDirective} & \texttt{omp target parallel for}\\
\textit{OMPTargetParallelForSimdDirective} & \texttt{omp target parallel for simd}\\
\textit{OMPTargetParallelGenericLoopDirective} & \texttt{omp target parallel loop} \\
\textit{OMPTargetSimdDirective} & \texttt{omp target simd} \\
\textit{OMPTargetTeamsDirective} & \texttt{omp target teams}\\
\textit{OMPTargetTeamsDistributeDirective} & \texttt{omp target teams distribute}\\
\textit{OMPTargetTeamsDistributeParallel- ForDirective} & \texttt{omp target teams distribute parallel for}\\
\textit{OMPTargetTeamsDistributeParallel- ForSimdDirective} & \texttt{omp target teams distribute parallel for simd}\\
\textit{OMPTargetTeamsDistributeSimdDirective} & \texttt{omp target teams distribute}\\
\textit{OMPTargetTeamsGenericLoopDirective} & \texttt{omp target teams loop}\\
\hline 
\end{tabular}
\label{tab:ASTKernelNodes}
\vspace{-3mm}
\end{table}
{\tt OMPDart} takes a compilable OpenMP program as input and relies on its AST for further analyses.
We use the LibTooling \cite{libtooling} library provided by Clang to build an AST for the input C/C++ code.
Operating on the AST presents a more complex challenge in assessing memory accesses in comparison to LLVM \textit{Intermediate Representation} (IR), but it helps maintain a closer relation to the source code. 
This relation makes analyzing ASTs more suitable for source-to-source transformations.

\begin{figure*}
\begin{minipage}{0.25\textwidth}
\begin{lstlisting}[
  language=C,
  tabsize=2,
  showspaces=false,
  showstringspaces=false,
  basicstyle=\ttfamily\small,
  breaklines=true,
  xleftmargin=20pt
]
int foo(int a[]) {
    int x = bar(a);
    if (x > 0) {
        a[x] = 0;
    }
    return x;
}
\end{lstlisting}
\end{minipage}
\qquad %
\begin{minipage}{0.65\textwidth}
\begin{tikzpicture}
[node distance=18ex and 10ex,->,>=stealth',thick,bend angle=15,auto,
    font=\footnotesize\vphantom{gl}, %
    edge from parent, 
    every node/.style={
    rectangle, minimum size=6mm, draw=black!75,
    thick, align=center},
    edge from parent/.style={draw=black!75,thick},
    level 1/.style={sibling distance=1.5cm},
    level 2/.style={sibling distance=1.0cm}, 
    level 3/.style={sibling distance=0.8cm}, 
    level distance=1.3cm,
    ]
    \node[rounded corners] (A) {Entry};
    \node[right of=A, rounded corners] (B) {Decl} 
        child { node {\texttt{int}}}
        child { node {$=$}
            child { node {\texttt{x}}}
            child { node {Call}
                child { node {\texttt{bar}}}
                child { node {\texttt{a}}}
            }
        }
    ;
    \node[right of=B, rounded corners] (C) {Pred} 
        child { node {$>$}
            child { node {\texttt{x}}}
            child { node {\texttt{0}}}
        }
    ;
    \node[right of=C, rounded corners] (D) {Stmt} 
        child { node {$=$}
            child { node {\texttt{a[x]}}
                child { node {\texttt{a}}}
                child { node {\texttt{x}}}
            }
            child { node {\texttt{0}}}
        }
    ;
    \node[right of=D, rounded corners] (E) {Exit} 
    ;

    \draw (A) edge[dotted] node[draw=none] {$\epsilon$} (B);
    \draw (B) edge[dotted] node[draw=none] {$\epsilon$} (C);
    \draw (C) edge[dotted] node[draw=none] {$true$} (D);
    \draw (D) edge[dotted] node[draw=none] {$\epsilon$} (E);
    \draw (C) edge[dotted,bend left=3.5em] node[draw=none] {$false$} (E);

    \draw ([yshift=1.5em] current bounding box.south west) -- ++(1cm,0) node[right,draw=none]{AST edge};
    \draw[dotted] ([yshift=0em] current bounding box.south west) -- ++(1cm,0) node[right,draw=none]{CFG edge};
\end{tikzpicture}
\end{minipage}
\caption{Example of AST-CFG representation. The graphical figure on the right is the AST-CFG representation of the code on the left.}
\label{fig:ast-cfg}
\vspace{-3mm}
\end{figure*}
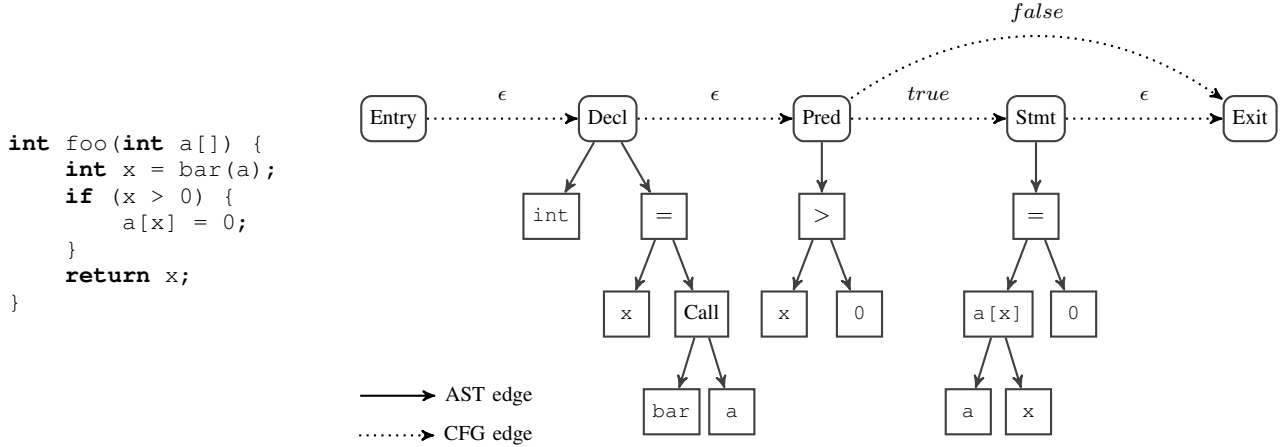

{\tt OMPDart} begins by parsing the AST to identify memory accesses associated with each variable reference. 
The memory accesses are grouped by parent function and classified as read, write, read/write, or unknown. 
There are many cases in which the type of memory access may be initially unknown, such as when a pointer is passed as a parameter to another function.

In certain scenarios, discerning memory accesses necessitates cross-translation unit analysis, particularly when a function definition resides in another translation unit.
Currently, our tool is confined to analyzing individual translation units, so conservative assumptions are derived from declared function prototypes.
We assume data passed via a pointer to \texttt{const} is strictly read-only, while non-static global variables and other parameters passed by pointer are handled conservatively.

In this stage, conditionals, loops, and target directives are identified and used to construct a CFG for each function definition in the input file. 
\texttt{IfStmt} and \texttt{SwitchStmt} AST nodes are classified as conditionals and \texttt{ForStmt}, \texttt{WhileStmt}, and \texttt{DoStmt} are AST nodes recognized as loops.
Nodes of the CFG are marked as to whether they are offloaded or not, depending on whether they belong to an offloaded region.
Offload kernels are blocks of code designated to be executed on a target device.
A list of \texttt{OMPExecutableDirective} AST nodes recognized as offload kernels can be seen in Table \ref{tab:ASTKernelNodes}.
This list includes all \texttt{target} directives except the \texttt{target (enter/exit) data} and \texttt{target update} directives.

The AST and CFG are combined to form a hybrid AST-CFG representation in which each node of the CFG is linked with the corresponding AST representation.
To the best of our knowledge, this is a novel data structure, and its application in this context is unique. The AST-CFG is integral to our analysis since it maintains both the structural hierarchy of the code represented by the AST and the control flow relationships of the CFG. 
A traversal of the CFG edges models data flow, variable liveness, and data validity to determine data dependencies, while intermittent AST analysis assists in detecting array access patterns that are crucial for deciding when to transfer data.
Figure \ref{fig:ast-cfg} shows example code and the corresponding AST-CFG representation.
The AST-CFG is constructed on a function level, similar to the process of constructing a \textit{Code Property Graph} (CPG).
CPGs were first introduced by Yamaguchi et al. \cite{yamaguchi2014} for the purposes of detecting vulnerabilities in C code.

\subsection{Interprocedural Analysis}

Now that each procedure has been identified and has a CFG associated with it, an interprocedural analysis step is applied. 
This step analyzes how function calls affect the data of the caller.
Specifically, we are interested in the passing of parameters and the use of common global variables across functions. 
The tool iteratively checks each function to determine if and in what manner any global variables or data at the address of pointer parameters were accessed. 
This includes whether data is read or written and if such accesses occur on the host or device. 
Once a determination is made about the use of pointer parameters or global variables, the model is augmented at each call site of the function with maximally pessimistic assumptions about the memory accesses of the callee. 
In this context, a maximally pessimistic approach means the analysis assumes the worst-case scenario for any unknown or uncertain information.
This conservative approach helps to ensure that we don’t violate the correctness of a program, of course, with the potential downside that we may miss an optimal solution.
This is particularly relevant in cases when we don’t have visibility into a function that may access memory in unknown ways.
This process can be repeated several times up to the maximum call depth of any function.
Each pass provides new information about the function to the model, but can be stopped early if no updates are made during a pass.
This helps the model make more accurate assumptions about data accesses.

Function inlining is an alternative approach that could be considered here. 
The main advantage would be the potential to avoid interprocedurally invalid paths.
However, applying inlining can cause CFG size to balloon exponentially. 
In addition, inlining cannot handle mutually recursive functions on its own.
Heuristics could be used to decide when to apply inlining to manage this trade-off between precision and CFG size. 
Even without employing an inlining approach, effective decisions can still be made reliably, as shown in section \ref{sec:results}.

\subsection{Data-Flow Analysis} \label{sec:data_flow_analysis}

The analysis is performed as a forward traversal of the AST-CFG for each function.
We trace the reads and writes to any variable referenced inside any offloaded region with the goal of identifying data dependencies between the host and device memory spaces.
Anti  (Write After Read, WAR) dependencies and output (Write After Write, WAW) dependencies do not necessitate any communication.
Only true  (Read After Write, RAW) dependencies need to be resolved by inserting one of the constructs described in Table \ref{tab:InsertableConstructs}.

\begin{table}
   \centering
   \caption{OpenMP Constructs That {\tt OMPDart} Will Insert to Resolve Data Dependencies}
   \begin{tabular}{p{.25\columnwidth}p{.65\columnwidth}}
      \toprule
      OpenMP Construct & Description \\
      \midrule
      \texttt{map(to:)} & on region entry copies data from host to device \\ \addlinespace[0.5em]
      \texttt{map(from:)} & on region exit copies data from device to host \\ \addlinespace[0.5em]
      \texttt{map(tofrom:)} & on region entry copies data from host to device and on exit copies data from device to host \\ \addlinespace[0.5em]
      \texttt{map(alloc:)} & on region entry allocates memory on device \\ \addlinespace[0.5em]
      \texttt{update to()} & updates data on device with the value from host \\ \addlinespace[0.5em]
      \texttt{update from()} & updates data on host with the value from device \\ \addlinespace[0.5em]
      \texttt{firstprivate()} & on region entry initializes a private copy on the device with the original value from the host \\
      \bottomrule
   \end{tabular}
   \label{tab:InsertableConstructs}
\end{table}

We define data to be \textit{valid} in a memory space if the data was last written to in said memory space and \textit{invalid} or \textit{stale} if the data was last written to in any other memory space.
While traversing the CFG, we track whether a memory space has a valid, up-to-date copy of each variable at each node.
If a node is reached that writes to the host environment, the corresponding data on the target is marked as stale.
If a node is reached that may read stale data, we have identified a true dependency and must evaluate the most efficient way to satisfy it.

For each function with at least one true dependency, we create a single \texttt{target data} region that encompasses all the kernels in the function's body.
The starting point of the region is determined by finding the start of the earliest offload kernel, and the end location is the end of the last offload kernel in the function.
However, doing this for a function consisting of a loop containing offload kernels would necessitate data being moved every iteration, possibly dominating execution time.
To avoid this and maximize the opportunity for data reuse, we must extend the \texttt{target data} region to begin before any loop capturing the first kernel and end after any loop capturing the last kernel.
Having a single \texttt{target data} region for each function enables general data mapping solutions since we can efficiently satisfy any dependencies between this data region and the rest of the function by using \texttt{map(to/from/tofrom:)} constructs and using a combination of \texttt{map(alloc:)} and \texttt{update} constructs to resolve dependencies between kernels and across loop iterations.
A single data region introduces the additional requirement that any variable declaration in the function body used by both the host and device must precede the location at which the tool intends the placement of the \texttt{target data} region.
If the input program violates this, the tool will detect this and issue an error indicating before which point the programmer should move the declaration so that the tool can continue successfully. 

Identifying the ideal insertion point for OpenMP data mapping constructs is crucial for minimizing data transfer, as Section \ref{sec:array_access_pattern_analysis} demonstrates.
Misplacing a data construct in a loop when it could be placed outside the loop body will almost definitely incur a significant performance penalty.
With the single \texttt{target data} region per function strategy that we have established, the difficulty of this problem becomes identifying correct and efficient placements of \texttt{update} constructs.

Upon encountring a branch in the CFG, each path is explored until they converge again or the function exits.
Nodes are only explored once since if we reach a visited node, we know we have found a loop and need to make any data valid as it was prior to the already visited node.
For instance, if data is valid on a device upon entering a loop, it should be valid at the end of the loop body so that each memory space is set up for the next iteration of the loop.
If a node that is part of a loop needs updated data from another environment, we make note until we reach the end of the loop.
If, by the end of the loop body, the data from the earlier node's environment is still valid, a data dependency does not exist between loop iterations, and we can safely map the data at a location prior to the loop.
Otherwise, there is a data dependency between loop iterations and an \texttt{update} statement must be inserted somewhere in the loop body to preserve correctness.

For some special cases, we can employ the \texttt{firstprivate} clause to improve efficiency over the \texttt{update to} construct.
The \texttt{firstprivate} clause creates private copies of variables for each thread and initializes them with the value from the master thread. 
When a \texttt{firstprivate} scalar variable is used within an OpenMP offload directive that launches a kernel, it may pass the initial value from the host as an argument.
This approach avoids additional memory allocations and copies when compared to using the \texttt{map(to:)} clause, as described in the Intel oneAPI GPU Optimization Guide \cite{intel2023}.
Therefore, it is advisable to use the \texttt{firstprivate} clause for transferring read-only scalar variables rather than the map clause. 
It's important to note that this argument passing behavior is not explicitly defined in the OpenMP 5.2 Specification \cite{openmpSpec52}, but is rather left to the discretion of the compiler implementation and may further depend on the architecture of the target accelerator. 
While the referenced Intel Optimization Guide \cite{intel2023} specifically addresses the icx implementation, we have additionally verified that this behavior applies to both clang 17.0.4 and gcc 13.2.1 on Nvidia GPUs. 
This can be verified with the NVIDIA NSight System profilers \cite{nvprof,nsys} to observe the reduction of host to device memcpys when using \texttt{firstprivate} over \texttt{map(to:)}. 
This is a specialized performance optimization that may not be widely known.
By using this tool, programmers can take advantage of such lesser-known performance optimizations.
When implementing this optimization, the analysis must be revised so that at the end of the kernel region, the model's internal state is updated to reflect that the \texttt{firstprivate} data is no longer valid.

Upon reaching the end of the target data region for a function, the problem becomes a liveness problem.
For variables used in an offloaded region, we want to determine if they are subsequently read, since if read after the target region we must make sure that data will be valid upon region exit.
If stale data may be read on the host after exit from a target data region, the \texttt{from} map-type is included in the variable's mapping clause to ensure the data will be valid.

\begin{lstlisting}[
  label=lst:for_bounds,
  caption={Sample C code consisting of a \texttt{for} loop with OpenMP target construct used to generate the AST in Listing \ref{lst:for_bounds_ast}.},
  captionpos=b,
  language=C,
  numbers=left,
  stepnumber=1,
  tabsize=2,
  showspaces=false,
  showstringspaces=false,
  basicstyle=\ttfamily\footnotesize,
  xleftmargin=20pt,
  float,  %
  belowskip=-\floatsep,
]
#define N 100
int main() {
    int a[N];
    ..
    #pragma omp target teams distribute \
            parallel for
    for (int i = 0; i < N/2; i++) {
        a[i] = i;
    }
    ..
}
\end{lstlisting}

\subsection{Array Access Pattern Analysis} \label{sec:array_access_pattern_analysis}

In \cite{guo2023}, Guo et al. presented a compile-time algorithm that operates on the Clang AST and can successfully eliminate the transfer of unused array segments in C and C++ OpenMP programs.  
The algorithm makes determinations about what data transfer is truly necessary based on the analysis of \texttt{for} loops to determine which array bounds are accessed. 
This analysis may be impeded if the programmer's use of a \texttt{for} statement lacks any of the initializing, conditional, or incrementing/decrementing statements, or if any of these statements are overly complex.
Furthermore, the algorithm presented in \cite{guo2023} has no mechanism to work with nested loops and uses the lifetime of a for loop as the lifetime to map the data.
In this section, we describe how we modify their approach for the purposes of arrays with multiple dimensions and nested loops, and how we determine efficient placement of update directives in scenarios involving nested loops.

\begin{lstlisting}[
  label=lst:for_bounds_ast,
  caption={Excerpt of the Clang AST dump generated from the source code in Listing \ref{lst:for_bounds} by running the command \texttt{clang -Xclang -ast-dump -fsyntax-only file.c}. Addresses, source locations, and other information has been omitted for readability, indicated by ``\texttt{..}''.},
  captionpos=b,
  numbers=left,
  stepnumber=1,
  showspaces=false,
  showstringspaces=false,
  basicstyle=\ttfamily\footnotesize,
  xleftmargin=20pt,
  float,  %
  aboveskip=-\parskip,
  belowskip=-\parskip,
]
ForStmt
|-DeclStmt ..
| `-VarDecl .. used i 'int' cinit
|   `-IntegerLiteral .. 'int' 0
|-<<<NULL>>>
|-BinaryOperator .. 'bool' '<' ..
| |-DeclRefExpr .. 'i' 'int'
| `-BinaryOperator .. 'int' '/'
|   |-IntegerLiteral .. 'int' 100
|   `-IntegerLiteral .. 'int' 2
|-UnaryOperator .. 'int' postfix '++'
| `-DeclRefExpr .. 'i' 'int'
`-CompoundStmt .. 
  `-BinaryOperator .. 'int' lvalue '='
    |-ArraySubscriptExpr ..
    | |-DeclRefExpr .. 'a' ..
    | `-DeclRefExpr .. 'i' ..
    `-DeclRefExpr .. 'i' 'int'
\end{lstlisting}

Listing \ref{lst:for_bounds_ast} shows the AST dump of the source code in Listing \ref{lst:for_bounds}. 
In Listing \ref{lst:for_bounds_ast}, each component of the \texttt{for} loop is recognizable. 
Lines 2 through 4 are the initialization statement which represents \texttt{int i = 0} from line 7 of the source code. 
We know 0 is the lower bound; since the iteration statement is a unary increment on \texttt{i} (AST lines 11-12) and the conditional statement has \texttt{i} on the left-hand side of the less than operator (AST lines 6-7). 
The upper bound can be found by evaluating the expression on the right-hand side of the less than operator, \texttt{100/2} (AST lines 6,8-10), and subtracting 1 to avoid an off-by-one error.

\begin{lstlisting}[
  label=lst:backprop_for,
  caption={Code snippet from the HeCBench's OpenMP offloading implementation of the Rodinia backprop benchmark \cite{hecbench} with a nested \texttt{for} loop which is executed on the host and calculates a sum.},
  captionpos=b,
  language=C++,
  numbers=left,
  stepnumber=1,
  tabsize=2,
  showspaces=false,
  showstringspaces=false,
  basicstyle=\ttfamily\footnotesize,
  xleftmargin=20pt,
  float=t,  %
  aboveskip=-\parskip,
  belowskip=-\floatsep,
]
// partial_sum: valid on device, invalid on host
for (int j = 1; j <= hid; j++) {
    sum = 0.0;
    for (int k = 0; k < num_blocks; k++) {
        sum += partial_sum[k * hid + j-1];
    }
    sum += input_weights[0][j];
    hidden_units[j] = 1.0 / (1.0 + exp(-sum));
}
\end{lstlisting}

{\tt OMPDart} extends this bounds analysis to infer the access patterns in nested loops to assist the placement of OpenMP \texttt{update} directives. 
Consider the nested \texttt{for} loop from HeCBench's OpenMP offloading implementation of the Rodinia backprop benchmark \cite{zheming2021} shown in Listing \ref{lst:backprop_for}. 
The code is executed on the host and reads from the array \texttt{partial\_sum}, which happens to be invalid on the host upon entering the snippet.
Using the logic presented in Section \ref{sec:data_flow_analysis}, the automated tool will correctly identify that somewhere preceding line 5, it must insert the following OpenMP directive: \texttt{target update from(partial\_sum)}. 
The immediate choice would be to insert the statement between lines 4 and 5, however, this would cause the data transfer to occur every iteration of the inner loop. 
To avoid this redundant data transfer, the \texttt{update} directive should be placed before the outermost loop (preceding line 2).
While both the options would produce technically correct code, in this example, placing the \texttt{update} directive in the inner loop would result in over $2$GB of data transfer compared to less than $5$MB when the directive is inserted before the \texttt{for} loops (with 65536 input elements).
By correctly identifying the nested loop structure and access pattern of \texttt{partial\_sum}, {\tt OMPDart} can make a better decision on the placement of the \texttt{update} directive, which translates to a 14$\times$ speedup.

\begin{algorithm}
\caption{Determining placement of \texttt{target update to/from()} OpenMP directives for array accesses in nested loops of arbitrary depth.}
\label{alg:findinsertloc}
\begin{algorithmic}
\Procedure{FindUpdateInsertLoc}{$a,loops,locLim$}
  \State $idxExpr \gets a.arraySubscript.idx$
  \State $indexingVars \gets getReferencedVars(idxExpr)$
  \State $pos \gets a$
  \While{$loops$ is not empty}
    \State $forStmt \gets loops.pop()$
    \If{$forStmt$ is before $locLim$ in file}
      \State \textbf{break}
    \EndIf
    \State $forIdxVar \gets findIndexingVar(forStmt)$
    \If{$forIdxVar$ is not a valid variable}
      \State \textbf{continue}
    \EndIf
    \If{$indexingVars.contains(forIdxVar)$}
      \State $pos \gets forStmt$
    \EndIf 
  \EndWhile
  \State \textbf{return} $pos$
\EndProcedure
\end{algorithmic}
\end{algorithm}

Algorithm \ref{alg:findinsertloc} can be used to find the outermost loop that impacts the indexing of a given array. 
$loops$ is a stack that contains references to \texttt{for} statements. 
The top of the stack corresponds to the \texttt{for} loop that includes array access $a$ in its body, with each subsequent element representing the enclosing loop. 
$locLim$ stores a statement which the directive must not precede, typically being the end of the preceding target kernel's scope. 
The procedure returns a pointer to the statement that the directive should directly precede or follow depending on whether the direction of data movement is \texttt{from} or \texttt{to} the device, respectively.
From there, this location is used to further decide where the best placement of the \texttt{update to/from} statement might be.

\subsection{Rewriter}
\label{sec:rewriter}
The primary task of the rewriter is to insert the data mapping directives/clauses determined by the model in the previous stage. 
Table \ref{tab:InsertableConstructs} shows the OpenMP constructs that the rewriter is responsible for inserting.
For this the rewriter must realize the exact insertion point in the source code corresponding to the indicated statement AST node from the model.

When only a single offload region exists in a function, and the beginning of the offload region is the insertion point for \texttt{target data} directive, the rewriter can simply append a \texttt{map} clause to the existing target directive. 
Otherwise, the rewriter will insert a new \texttt{target data} directive and increase the indentation of the captured block.
\texttt{firstprivate} clauses are simply appended to an existing OpenMP directive indicated by the model.
Determining the exact insertion point in the source code for \texttt{target update} directives requires additional work. 
The previous stage provides a pointer to a \texttt{Stmt} node, which must be preceded/followed by a \texttt{target update} directive, however, this may or may not be a semi-terminated statement. 
The logic varies slightly when determining the insertion point for an \texttt{update to} directive compared to an \texttt{update from} directive. 
Generally, \texttt{update from} directives should be inserted before the statement indicated by the model from the previous stage and after for \texttt{update to} directives. 
Special considerations are made for statements contained in a loop conditional. 
\texttt{update to} directives pertaining to a memory access in a loop conditional should be placed at the beginning of the loop body. 
\texttt{update from} directives for \texttt{for}/\texttt{while} conditionals may need to be placed either before the loop or immediately preceding the end of the loop body. 
Since \texttt{do} conditionals occur at the end of the loop, \texttt{update from} directives will always be inserted at the end of the loop body.
Finally, prior to inserting the directives and clauses into the source code, each type of directive and clause is consolidated based on their insertion point.
This process condenses the constructs into a directive per insertion point.
Finally, the directives are inserted into the source code at the determined locations.

\section{Benchmarks} \label{sec:benchmarks}

\begin{table*}
\centering
\caption{Programs used for evaluating {\tt OMPDart}.}
\begin{tabular}{p{0.22\linewidth}p{0.08\linewidth}p{0.12\linewidth}p{0.49\linewidth}}
\hline
\textbf{Application Name} & \textbf{Benchmark Suite} & \textbf{Domain} & \textbf{Description}\\
\hline
Accuracy of Prediction (accuracy) & HeCBench & Machine Learning & Computes the classification accuracy of a neural network \cite{pytorch}\\
Allen-Cahn Equation (ace) & HeCBench & Fluid Dynamics & Phase-field simulation of dendritic solidification \cite{allencahn}\\
Back Propagation (backprop) & Rodinia & Pattern Recognition & Machine learning algorithm that trains the weights of connecting nodes on a neural network \cite{mitchell1997}\\
Breadth First Search (bfs) & Rodinia & Graph Traversal & Traverses all the connected components in a graph \cite{harish2007}\\
Coulombic Potential (clenergy) & HeCBench & Physics Simulation & Evaluates electrostatic potentials on a 3-D lattice using direct Coulomb summation method \cite{clenergy}\\
HotSpot (hotspot) & Rodinia & Physics Simulation & Thermal simulation tool used for estimating processor temperature based on an architectural floor plan and simulated power measurements \cite{hotspot2006}\\
LULESH (lulesh) & HeCBench & Hydrodynamics & Proxy application that simulates shock hydrodynamics \cite{lulesh}\\
Needleman-Wunsch (nw) & Rodinia & Bioinformatics & Non-linear global optimization method for DNA sequence alignments \cite{needleman1970}\\
XSBench (xsbench) & HeCBench & Neutron Transport & Mini-app representing a key computational kernel of the Monte-Carlo neutron transport algorithm \cite{xsbench}\\
\hline 
\end{tabular}
\label{tab:benchmarks_desc}
\vspace{-3mm}
\end{table*}
To evaluate {\tt OMPDart}, \textit{nine} OpenMP GPU programs/benchmarks were studied. Four benchmarks were chosen from the Rodinia GPU suite \cite{rodinia-gpu}, and four from the HeCBench suite \cite{hecbench}, detailed in Table \ref{tab:benchmarks_desc}.
The selection aimed to cover a range of complexities, with a preference for benchmarks featuring a mix of multiple kernels, numerous mapped variables, several \texttt{map} clauses, and the inclusion of \texttt{update} clauses. 

Table \ref{tab:benchmark_complexity} details the complexities of each of the selected benchmarks, including the number of kernel regions, lines of code in offloaded regions, number of mapped variables, and an approximation of the number of possible mappings.
The number of possible mappings is approximated by the sum of two parts.
(1) The total combinations of mapping clauses. Each variable in each kernel have one of the $\textbf{4}$ map-type types in \{\texttt{to}, \texttt{from}, \texttt{tofrom}, \texttt{alloc}\}.
(2) The total combinations of \texttt{update} clauses. \texttt{update} clauses can be placed between every statement and there are $\textbf{3}$ options for each variable which are \texttt{to}, \texttt{from}, or not to insert an \texttt{update} directive at this location.
We divide the number of lines by $\textbf{2}$ since this provides a more conservative representation of the actual number of effective placements when taking into account that some statements span multiple lines, some lines are empty, and some only contain comments.
The number of possible mappings is calculated as follows:
\[
\textnormal{mappings} = \textnormal{kernels} \times \textnormal{variables} \times 4 + \bigg\lfloor\frac{\textnormal{lines}}{2}\bigg\rfloor \times \textnormal{variables} \times 3 
\]

\begin{table}
   \centering
   \caption{Comparison of Benchmark Data Mapping Complexity}
   \begin{tabular}{lrrrr}
      \toprule
      Benchmark  &
      Kernels &
      \begin{tabular}{@{}l@{}}Offloaded \\ Lines\end{tabular} &
      \begin{tabular}{@{}l@{}}Mapped \\ Variables\end{tabular} &
      \begin{tabular}{@{}l@{}}Possible \\ Mappings\end{tabular} \\
      \midrule
      \textbf{accuracy}  &  1 &   37 &  5 &    297 \\
      \textbf{ace}       &  6 &  202 &  7 &   2289 \\
      \textbf{backprop}  &  2 &   87 &  7 &    969 \\
      \textbf{bfs}       &  2 &   40 &  8 &    544 \\
      \textbf{clenergy}  &  2 &  103 &  5 &    812 \\
      \textbf{hotspot}   &  1 &   85 & 15 &   1972 \\
      \textbf{lulesh}    & 15 & 1293 & 65 & 129967 \\
      \textbf{nw}        &  2 &  122 & 12 &   2292 \\
      \textbf{xsbench}   &  1 &  271 &  8 &   3284 \\
      \bottomrule
   \end{tabular}
   \label{tab:benchmark_complexity}
\end{table}

To evaluate {\tt OMPDart}, three versions of each of the selected benchmark applications were evaluated.

\begin{enumerate}
    \item \textit{Unoptimized} -- This version is devoid of any explicit data mappings and falls back on the default compiler-defined handling of data transfer and reuse.
    \item \textit{OMPDart Optimized} -- The unoptimized version of each program was fed to {\tt OMPDart}, and the output is the tool’s optimized version of the program.
    \item \textit{Expert Optimized} -- This is a version of the applications as implemented in HPC benchmark suites Rodinia and HeCBench. All applications used for evaluation had explicit data mappings in them to help improve bottlenecks with data transfer and synchronization.
\end{enumerate}

\section{Experiments} \label{sec:results}

\begin{figure*}
    \centering
    \includegraphics[ width=0.9\textwidth]{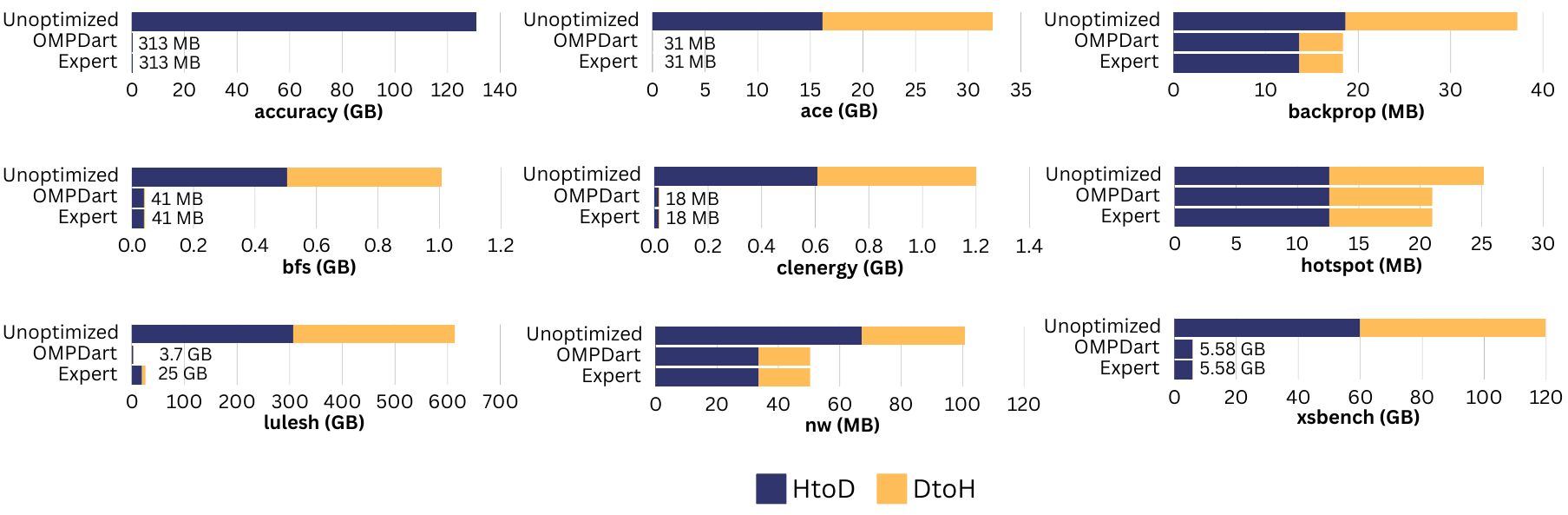}
    \caption{Comparison of GPU data transfer activity (bytes) (lower is better). \textbf{HtoD}: Data transfer from CPU to GPU. \textbf{DtoH}: Data transfer from GPU to CPU.}
    \label{fig:resultsTotalDataTransferred}
    \vspace{-3mm}
\end{figure*}
\begin{figure*}
    \centering
    \includegraphics[ width=0.9\textwidth]{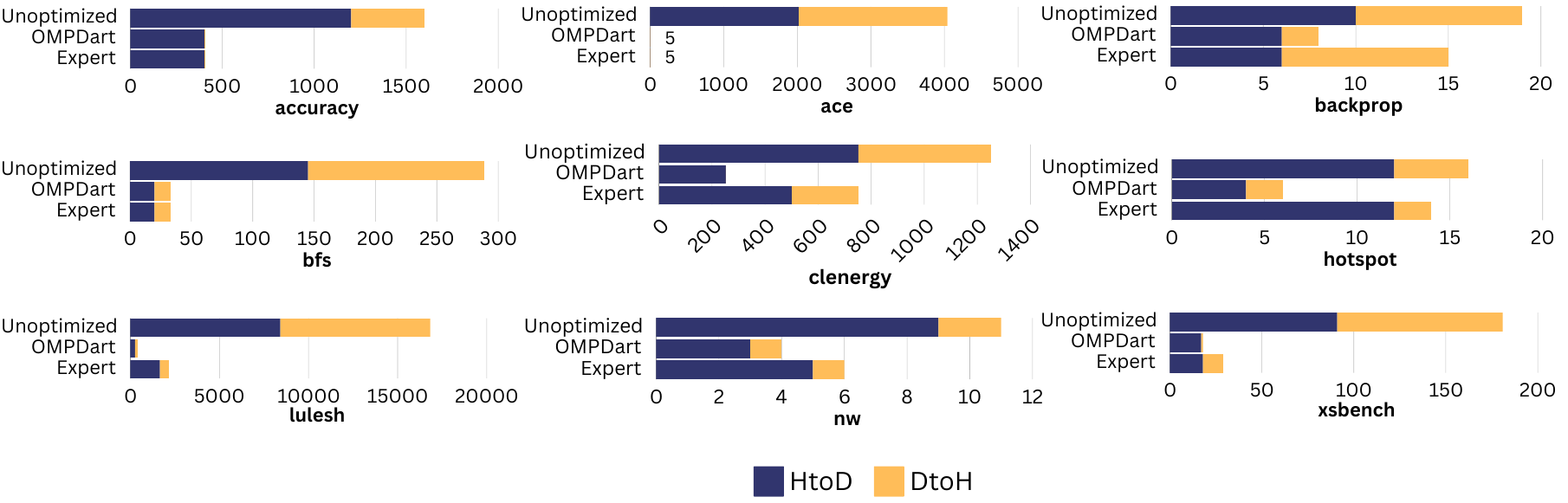}
    \caption{Comparison of GPU data transfer activity (\# calls) (lower is better). \textbf{HtoD}: Data transfer from CPU to GPU. \textbf{DtoH}: Data transfer from GPU to CPU.}
    \label{fig:resultsGpuActivity}
    \vspace{-3mm}
\end{figure*}

This section highlights our experimental evaluations of {\tt OMPDart} and compares its outputs with the unoptimized OpenMP offload code and the expert-implemented versions with explicit data mappings (Section \ref{sec:benchmarks}).

\textit{\textbf{Setup}.} We first execute and evaluate metrics for the unoptimized and expert versions of the OpenMP target offload codes.
Next, we pass the unoptimized application through {\tt OMPDart} to output data mappings that substantially reduce the data transfer overhead. Our results were obtained on an NVIDIA A100 running CUDA 11.8.89 compiled with Clang 17.0.4.

\textit{\textbf{Results and Analysis}.}
\begin{figure}
    \centering
    \includegraphics[ width=0.4\textwidth]{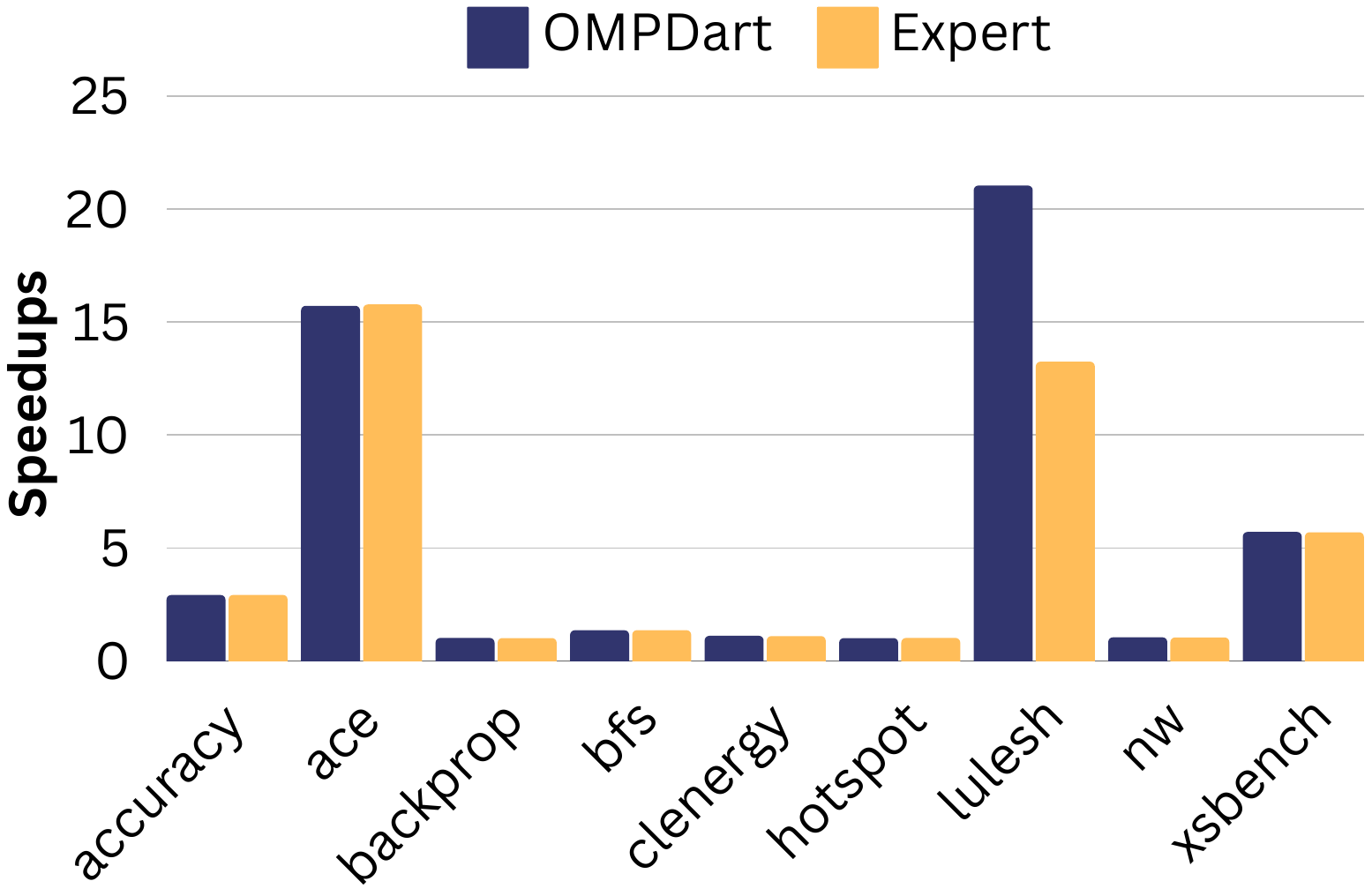}
    \caption{Speedups over unoptimized OpenMP offload code. (Higher is better)}
    \label{fig:resultsExecTime}
\end{figure}
\begin{figure}
    \centering
    \includegraphics[ width=0.4\textwidth]{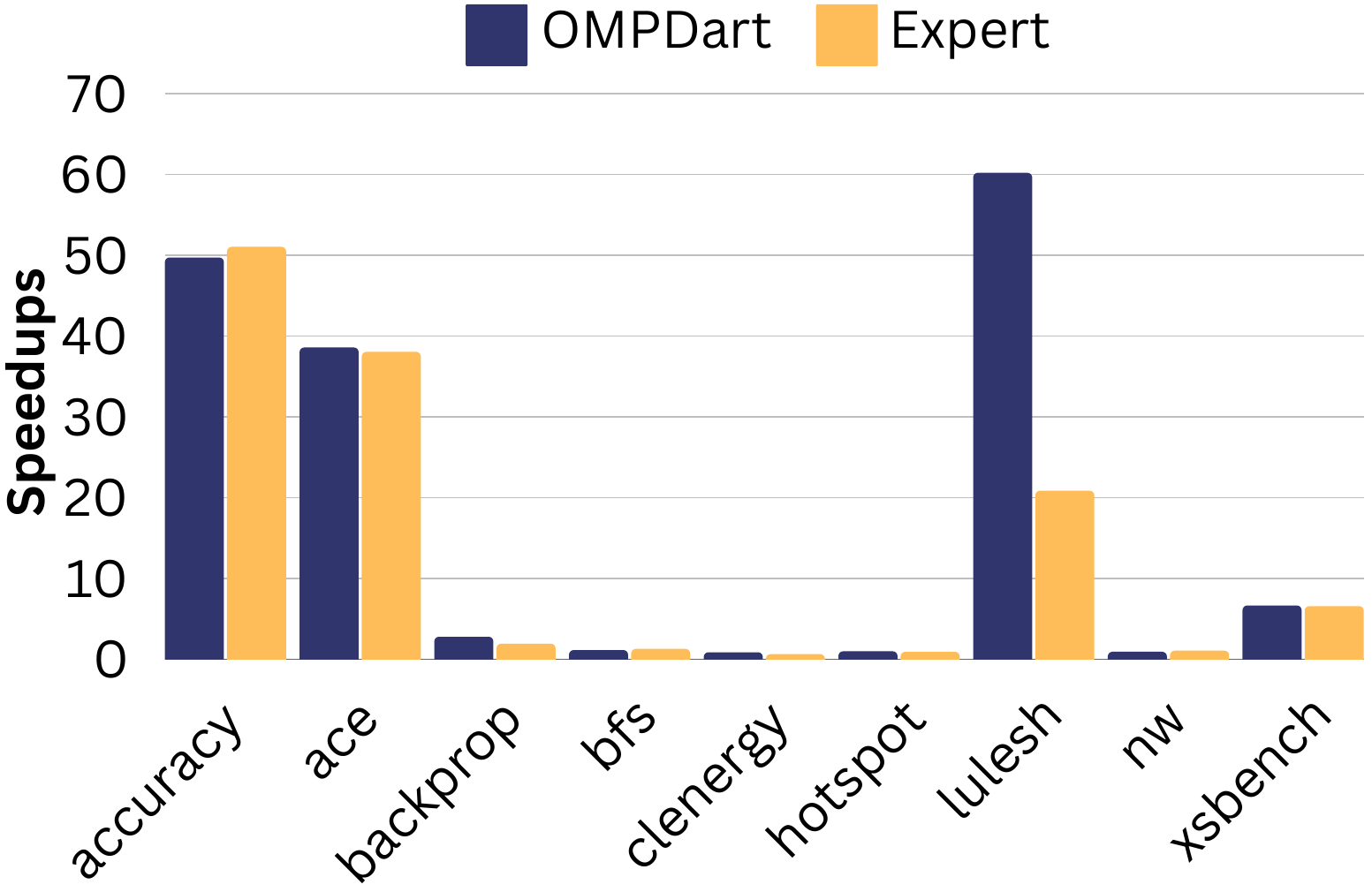}
    \caption{Improvements in data transfer wall times over unoptimized data mappings. (Higher is better)}
    \label{fig:dataTransferTime}
    \vspace{-3mm}
\end{figure}

The OpenMP applications with data mappings generated by {\tt OMPDart} were compiled with the same options as used for the unoptimized and expert implementations.
The compiled application was then executed to collect the runtimes.
Each application was also profiled with NVIDIA Nsight Systems (\textit{nsys}) \cite{nsys} to evaluate the number of \textbf{H}ost-\textbf{to}-\textbf{D}evice (HtoD) and \textbf{D}evice-\textbf{to}-\textbf{H}ost (DtoH) CUDA \texttt{memcpy} calls, the bytes transferred each way, and the total time taken by data transfer.

The data mappings generated by {\tt OMPDart} significantly reduce the data transferred for all applications over the unoptimized versions as can be seen in Figure \ref{fig:resultsTotalDataTransferred}.
Figure \ref{fig:resultsGpuActivity} shows that {\tt OMPDart} successfully reduced GPU data transfer activity in terms of CUDA {\tt memcpy} calls below the level of the expert mappings in 6 of the benchmarks.
Figure \ref{fig:resultsExecTime} shows the speedup of each benchmark application with respect to the unoptimized implementation.
For each application, the mappings were always \textit{at least} as good as the expert implementations.

For {\tt ace}, {\tt accuracy}, and {\tt backprop}, {\tt OMPDart} produced data mappings identical to the expert-optimized program.
For {\tt clenergy}, the tool-generated mappings were equivalent to the expert mappings except for one additional mapping for a struct that the expert presumably had overlooked.
This translates to a drastic 66\% reduction in CUDA \texttt{memcpy} calls, but since the size of the struct is relatively small, there is only a reduction of 60kB in total data transfer and a marginal $1.01\times$ speedup over the expert implementation.
The mappings generated for {\tt bfs} differed from those of the expert due to the tool opting to resolve data dependencies through separate \texttt{update to} and \texttt{update from} clauses, whereas the expert version employed a single \texttt{map} clause to achieve an equivalent outcome.
The tool employed the use of the \texttt{firstprivate} clause to reduce the number of CUDA {\tt memcpy} calls compared to the expert-optimized program in {\tt hotspot}, {\tt nw}, and {\tt xsbench}.
As can be observed in Figure \ref{fig:resultsGpuActivity}, this amounts to a reduction in CUDA {\tt memcpy} calls by $57\%$ in {\tt hotspot}, $33\%$ in {\tt nw}, and $38\%$ in {\tt xsbench} over the expert defined mappings.

Across these benchmarks, {\tt OMPDart} reduces the data transferred between host and device by $1010\times$, $400\times$, $2\times$, $65\times$, $23\times$, $1.2\times$, $2\times$, and $20\times$ over default OpenMP offload data mappings for {\tt ace}, {\tt accuracy}, {\tt backprop}, {\tt clenergy}, {\tt bfs}, {\tt hotspot}, {\tt nw}, and {\tt xsbench}.
{\tt OMPDart} helps improve runtime performance by $16\times$, $2.9\times$, $1.01\times$, $1.11\times$, $1.36\times$, $1.01\times$, $1.04\times$, and $5.7\times$ for these applications respectively.

{\tt OMPDart} generated mappings significantly outperformed the expert-defined mappings in {\tt lulesh}, achieving a speedup of 1.6$\times$ and a reduction in data transfer of over 23GB.
This improvement can be primarily attributed to the inclusion of several redundant \texttt{update} directives in the expert implementation.
{\tt OMPDart} helps remove such redundant mappings, which reduces the data transfer bottleneck and decreases synchronization overheads.
Data mapping selections made by {\tt OMPDart} help reduce host-to-device (HtoD) data transfers by $7.4\times$ over expert implementation and help reduce device-to-host (DtoH) data transfers by $5.1\times$.
{\tt lulesh} uses $65$ variables across the host environment and $15$ kernels over thousands of lines of code.
Manually tracing their use and identifying the best implementation is an extremely time-consuming and challenging task.
Such complex applications have high potential for programmers to unintentionally create redundant data transfer, especially since compilers do not generate warnings for this.
Our tool can go a long way in trying to help address such challenges.

As mentioned in Section \ref{sec:input}, programmers might sometimes deliberately use stale data.
The only benchmark studied that relies on the use of stale data is {\tt backprop}.
After running the tool on {\tt backprop} an additional update directive was added by hand to adhere to this program behavior. 
All other mappings were generated automatically.

The tool’s execution time is negligible. The average execution time of {\tt OMPDart} on the benchmarks studied was $0.29$s.
{\tt lulesh}, with 15 kernels, was the biggest benchmark evaluated and had the greatest overhead of $1.35$s.
Table \ref{tab:tool_overhead} lists the tool’s execution time to analyze and transform each benchmark.

\begin{table}
   \centering
   \caption{{\tt OMPDart} Overhead}
   \begin{tabular}{lr}
      \toprule
      Benchmark  &
      Tool Execution Time \\
      \midrule
      \textbf{accuracy}  &  0.34s \\
      \textbf{ace}       &  0.18s \\
      \textbf{backprop}  &  0.34s \\
      \textbf{bfs}       &  0.14s \\
      \textbf{clenergy}  &  0.16s \\
      \textbf{hotspot}   &  0.05s \\
      \textbf{lulesh}    &  1.35s \\
      \textbf{nw}        &  0.14s \\
      \textbf{xsbench}   &  0.14s \\
      \bottomrule
   \end{tabular}
   \label{tab:tool_overhead}
\end{table}

Overall, across all studied benchmark applications, {\tt OMPDart} generated mappings produce a geometric mean speedup of 2.8$\times$ over the default implicit data mapping rules and 1.05$\times$ speedup over the expert-defined mappings.
A geometric mean reduction of 2.1GB in data transfer is saved by the tool over the default implicit data mapping rules.
These results further affirm the critical importance of defining efficient data mappings.
Furthermore, our evaluations show the impact that data transfer has on the overall application performance.
To demonstrate this, we have also analyzed the wall times spent by each application on data transfer and synchronization.
Almost all applications are better than the unoptimized implementation when {\tt OMPDart} and expert mappings are evaluated on this metric.
As seen in Figure \ref{fig:dataTransferTime}, time taken for data transfer improves by up to $60\times$ and across all applications improves by a geometric mean of $5.1\times$ when {\tt OMPDart} mappings are used.
The improvements in time taken to transfer data and synchronize are similar in most cases for the expert implementation.
Only in {\tt lulesh} do the expert mappings perform significantly worse on this metric compared to {\tt OMPDart}, which offers improvements in data transfer time of $20\times$.
Overall, for all applications the geometric mean improvements for the expert implementation is $4.2\times$.

Each program with the mappings generated by {\tt OMPDart} was also evaluated for correctness by comparing its output to the output produced by the expert implementation.
The mappings generated by our tool produced outputs consistent with those produced by experts.

\section{Discussion} \label{sec:discussion}

{\tt OMPDart} is a source-to-source transformation tool which makes these optimizations compiler agnostic.
OpenMP offloading support and performance varies drastically between compilers and architectures \cite{davis2021, diaz2019, thavappiragasam2022, openmpCompilers}.
Some compilers, oftentimes hardware vendor-specific compilers, only support a small set of architectures but incorporate many architecture-specific optimizations that may make a compiler the best option for offloading to certain targets.
By implementing general data mapping optimizations on a source-code level we ensure portability and enable the programmer to use the compiler best suited to the target system.

Static analysis faces inherent limitations due to foundational results in computability theory, notably Rice's Theorem \cite{rice1953} and Turing's introduction of the Halting Problem \cite{turing1937}.
The implications for static analysis necessitate conservative assumptions and approximations, which for static data-flow analyses means balancing precision and soundness.
Our analysis errs on the side of soundness.

We assume that pointers can be disambiguated through alias analysis.
If alias analysis fails to determine whether two pointers in a program can refer to the same memory location, the analysis will fail.
Moreover, when there is a single element access of an array it is conservatively assumed to be the access of the entire array.
This can be a source of correctness violation and can be mitigated by more precisely tracking accesses for array elements and using more sophisticated array access bounds analysis.

When using {\tt OMPDart} it should be noted that it is currently limited to analyzing a single translation unit at a time.
It performs best when it has full visibility into all functions.
Functions defined in external translation units require conservative, worst-case assumptions about the behavior of those functions.
In the future, the interprocedural analysis could be expanded to enable improved analysis capabilities for programs organized with kernels in separate translation units.
The array bounds analysis presented in Section \ref{sec:array_access_pattern_analysis} is currently only implemented for \texttt{for} loops, however, in the future this logic could be extended to encompass \texttt{while} and \texttt{do} loops.
One approach for doing this would be to analyze the code preceding the loop to determine initial conditions, using the conditional statement of the \texttt{while}/\texttt{do} statement, and analyzing the loop body for possible incrementing statements.

At the time of our evaluation, AMD GPUs were not readily available to us, hence they were not tested.
However, we believe a similar reduction in data transfers should be seen on AMD GPUs as well.

\section{Related Work} \label{sec:related_work}

Mishra et al. \cite{mishra2020}, presented a tool that uses the Clang AST to identify nearby OpenMP offload kernels and create data mappings across multiple kernels.
The program analysis presented here did not consider data dependencies between kernel regions or loop iterations and cannot insert \texttt{update} or \texttt{firstprivate} constructs.
Unlike {\tt OMPDart}, this study only considered data mappings within the same function body.

Another work, OpenMP Advisor \cite{mishra2023}, utilizes machine learning to predict performant OpenMP offloading directives.
As part of this work, the authors used a Data Transfer and Reuse Analysis Tool they had previously developed to predict the data mappings for programs before running it through the machine learning model that they developed.

Recently, Guo et al. \cite{guo2023} developed compile-time data transfer optimizations that center around filtering the transfer of unused array segments.
They accomplish this automatically at compile time-based on analysis of the Clang AST.
Guo et al. present techniques for array-bound analysis based on loop structures which we have used and extended in Section \ref{sec:array_access_pattern_analysis}.

OMPSan (OpenMP Sanitizer) \cite{barua2019} is a static analysis tool that verifies the correctness of and identifies bugs related to OpenMP data mapping constructs.
OmpMemOpt \cite{barua2020} employs a compile-time analysis that aims to identify and eliminate redundant data movements.
This is accomplished by formulating a partial redundancy elimination problem and applying the lazy code motion compiler optimization technique.
{\tt OMPDart} differs from OmpMemOpt in that we focus on generating efficient data mappings while OmpMemOpt offers compiler optimizations to automate the removal of redundant memory copies at compile time.
Due to the unavailability of OmpMemOpt, direct performance comparisons could not be made.

\section{Conclusion} \label{sec:conclusion}

In this work, we addressed the time-consuming and error-prone task of manually defining efficient OpenMP data mappings.
The efficient movement of data is necessary for maximizing the utilization of modern heterogeneous computing platforms.
We proposed a static data-flow analysis for C and C++ source code that can correctly generate performant data mappings and automatically insert the data directives into OpenMP source code.
The tool was evaluated on 9 benchmarks and compared against expert implementations from the Rodinia and HeCBench suites.
Our results show a geometric mean speedup of 2.8$\times$ over the default implicit data mapping rules and 1.05$\times$ speedup over the expert-defined mappings.
Across all 9 benchmarks, our tool also demonstrated a substantial reduction of redundant data transfer, with a geometric mean reduction of 2.1GB.
These findings highlight the effectiveness of our approach in automatically generating performant data mappings for OpenMP offloading applications.
By automating this process, our tool not only enhances the performance of OpenMP offloading applications but also saves programmers significant time and effort, accelerating the development of such applications.

\section*{Acknowledgements}

This research was supported by the National Science Foundation under Grant number 2211982.
We would also like to thank the Research IT team\footnote{\url{https://researchit.las.iastate.edu/}} of Iowa State University for their support and for providing access to the HPC Cluster, which were essential for conducting the experiments in this project.

\bibliographystyle{IEEEtran}
\bibliography{references}
\end{document}